\documentclass[galley,jgrga]{agu2001}
\usepackage{graphicx}
\authorrunninghead{WIEGELMANN}

\titlerunninghead{Coronal magnetic fields.}

\authoraddr{Max-Planck-Institut f\"ur Sonnensystemforschung,
Max-Planck-Strasse 2, 37191 Katlenburg-Lindau, Germany,
(wiegelmann@mps.mpg.de)}

\begin{document}

%
%
%
%
%

%
%

\title{Nonlinear force-free modeling of the solar coronal magnetic field}
\author{T. Wiegelmann}
\affil{Max-Planck-Institut f\"ur Sonnensystemforschung,
Max-Planck-Strasse 2, 37191 Katlenburg-Lindau, Germany}

\begin{abstract}
The coronal magnetic field is an important quantity because the magnetic field
dominates the structure of the solar corona. Unfortunately direct measurements of
coronal magnetic fields are usually not available. The photospheric magnetic field
is measured routinely with vector magnetographs. These photospheric measurements are
extrapolated into the solar corona. The extrapolated coronal magnetic field depends
on assumptions regarding the coronal plasma, e.g. force-freeness. Force-free means
that all non-magnetic forces like pressure gradients and gravity are neglected. This
approach is well justified in the solar corona due to the low plasma beta. One has
to take care, however, about ambiguities, noise and non-magnetic forces in the
photosphere, where the magnetic field vector is measured. Here we review different
numerical methods for a nonlinear force-free coronal magnetic field extrapolation:
Grad-Rubin codes, upward integration method, MHD-relaxation, optimization and the
boundary element approach. We briefly discuss the main features of the different
methods and concentrate mainly on recently developed new codes.
\end{abstract}
%
%

%
\begin{article}
%
%
%
\section*{Contents}
\begin{enumerate}
\item Introduction
\begin{enumerate}
\item [1.1.]Why do we need nonlinear force-free fields?
\end{enumerate}
\item Nonlinear force-free codes
\begin{enumerate}
\item [2.1.]Grad-Rubin method
\item [2.2.]Upward integration method
\item [2.3.]MHD relaxation
\item [2.4.]Optimization approach
\item [2.5.]Boundary element or Greens function like method
\end{enumerate}
\item How to deal with non-force-free boundaries and noise ?
\begin{enumerate}
\item [3.1.]Consistency check of vector magnetograms
\item [3.2.]Preprocessing
\end{enumerate}
\item Code testing and code comparisons
\begin{enumerate}
\item [4.1.]The NLFFF-consortium
\end{enumerate}
\item Conclusions and Outlook
\end{enumerate}
$ $ \\
\section{Introduction} $ $\\
Information regarding the coronal magnetic field are important for space weather
application like the onset of flares and coronal mass ejections (CMEs).
Unfortunately we usually cannot measure the coronal magnetic field directly,
 although recently some progress has been made
 \citep[see e.g.,][]{judge98,solanki:etal03,lin:etal04}. Due to the optically thin coronal
plasma direct measurements of the coronal magnetic field have a line-of-sight
integrated character and to derive the accurate 3D structure of the coronal magnetic
field a vector tomographic inversion is required. Corresponding feasibility studies
based on coronal Zeeman and Hanle effect measurements have recently been done by
\cite{kramar:etal06} and \cite{kramar:etal07}. These direct measurements are only
available for a few individual cases and usually one has to extrapolate the coronal
magnetic field from photospheric magnetic measurements. To do so, one has to make
assumptions regarding the coronal plasma. It is helpful that the low solar corona is
strongly dominated by the coronal magnetic field and the magnetic pressure is orders
of magnitudes higher than the plasma pressure. The quotient of plasma pressure $p$
and magnetic pressure, $B^2/(2 \mu_0)$    is small compared to unity
$(\beta=2 \mu_0 p/B^2 \ll 1)$. In lowest order non-magnetic forces like pressure
gradient and gravity can
be neglected which leads to the force-free assumption. Force free fields are
characterized by the equations:

\begin{eqnarray}
{\bf j}\times{\bf B}  & = & {\bf 0}        \label{forcebal}\\
\nabla \times {\bf B }& = & \mu_0 {\bf j}  \label{ampere}, \\
\nabla\cdot{\bf B}    & = &         0      \label{solenoidal},
\end{eqnarray}
where ${\bf B}$ is the magnetic field, ${\bf j}$ the electric current density and
 $\mu_0$ the permeability of vacuum.
 Equation (\ref{forcebal}) implies that for force-free fields the current
density and the magnetic field are parallel, i.e.

\begin{equation}
\mu_0 {\bf j} = \alpha {\bf B},
\label{jparb}
\end{equation}

or by replacing ${\bf j}$ with Eq. (\ref{ampere})

\begin{equation}
\nabla \times {\bf B }  =  \alpha {\bf  B} \label{amperealpha},
\end{equation}

 where $\alpha$ is called the force-free function.
To get some insights in the structure of the space dependent function $\alpha$,
 we take the divergence of Eq. (\ref{jparb}) and
 make use of Eqs. (\ref{ampere}) and (\ref{solenoidal}):

\begin{equation}
{\bf B} \cdot \nabla \alpha  = 0, \label{alphaeq}
\end{equation}

which tells us that the force-free function $\alpha$ is constant on every field
line, but will usually change from one field line to another. This
 generic case is called nonlinear force-free approach.

 Popular simplifications are $\alpha=0$ (current free potential fields,
 see e.g., \cite{schmidt64,semel67,schatten69,sakurai82})
 and $\alpha={\rm constant}$ (linear force-free approach, see e.g.,
 \cite{nakagawa:etal72,chiu:etal77,seehafer78,alissandrakis81,seehafer82,semel88}).
 These simplified models have been in particular
 popular due to their relative mathematical simplicity and because only
 line-of-sight photospheric magnetic field measurements are required.
 Linear force-free fields still contain one free global parameter
 $\alpha$, which can be derived by comparing coronal images with projections
 of magnetic field lines (e.g., \cite{carcedo:etal03}).
 It is also possible to derive an averaged value of $\alpha$
 from transverse photospheric magnetic field measurements
 (e.g. \cite{pevtsov:etal94,wheatland99,leka:etal99}).
 Despite the popularity and frequent use of these simplified models in the past,
 there are several limitations in these models (see below) which  ask for considering
 the more sophisticated nonlinear force-free approach.

 Our aim is to review recent developments of the extrapolation
 of nonlinear force-free fields (NLFFF). For earlier reviews on force-free
 fields we refer to
 \citep{sakurai89,aly89,amari:etal97,mcclymont:etal97}
 and chapter 5 of \cite{aschwanden:book}.
  Here we will concentrate mainly
 on new developments which took place after these earlier reviews. Our main
 emphasis is to study methods which extrapolate the coronal magnetic field
 from photospheric vector magnetograms. Several vector magnetographs are currently
 operating or planed for the nearest future, e.g., ground based: the solar flare
 telescope/NAOJ \citep{sakurai:etal95}, the imaging vector magnetograph/MEES Observatory
 \citep{mickey:etal96}, Big Bear Solar Observatory, Infrared Polarimeter VTT,
 SOLIS/NSO  \citep{henney06} and space born:
 Hinode/SOT \citep{shimizu04}, SDO/HMI \citep{borrero:etal06}.
 Measurements from these vector magnetograms will provide us eventually
 with the magnetic field
 vector on the photosphere, say $B_{z0}$ for the normal and $B_{x0}$ and $B_{y0}$ for
 the transverse field. Deriving these quantities from the measurements is an
 involved physical process, which includes measurements based on the
 Zeeman and Hanle effect, the inversion of Stokes profiles
 (e.g., \cite{labonte:etal99}) and removing the $180^\circ$ ambiguity
 (e.g., \cite{metcalf94,metcalf:etal06}) of the horizontal magnetic
 field component.
 Special care has to be taken for vector magnetograph measurements
 which are not close to the solar disk, when the line-of-sight and normal
 magnetic field component are far apart (e.g. \cite{gary:etal90}). For the purpose
 of this paper we do not address the observational methods and recent developments
 and problems related to deriving the photospheric magnetic field vector. We
 rather will concentrate on how to use the photospheric
 $B_{x0}, \, B_{y0}$ and $B_{z0}$ to derive the coronal magnetic field.

 The transverse photospheric magnetic field $(B_{x0}, \, B_{y0})$ can
 be used to approximate the normal electric current distribution by

 \begin{equation}
 \mu_0 j_{z0}=\frac{\partial B_{y0}}{\partial x}-\frac{\partial B_{x0}}{\partial y}
 \label{j_photo}
 \end{equation}

 and from this one gets the distribution of $\alpha$ on the photosphere by

 \begin{equation}
 \alpha(x,y)=\mu_0 \frac{j_{z0}}{B_{z0}}
 \label{alpha0direct}
 \end{equation}

 By using Eq. (\ref{alpha0direct}) one has to keep in mind that rather large uncertainties
 in the transverse field component and  numerical derivations used in
 (\ref{j_photo}) can cumulate  in significant errors for the current density. The problem
 becomes even more severe by using (\ref{alpha0direct}) to compute $\alpha$ in regions
 with a low normal magnetic field strength $B_{z0}$. Special care has to be
 taken at photospheric polarity inversion lines, i.e. lines along which $B_{z0}=0$
 \citep[see e.g.,][]{cuperman:etal91}.
 The nonlinear force-free coronal magnetic field extrapolation is a
 boundary value problem.
 As we will see later some of the NLFFF-codes
 make use of (\ref{alpha0direct}) to specify the boundary conditions
 while other methods use the photospheric magnetic field vector
 more directly to extrapolate the field into the corona.

 Pure mathematical investigations of the nonlinear force-free equations
 \citep[see e.g.][]{aly84,boulmezaoud:etal00,aly05}
 and modelling approaches not based on
 vector magnetograms are important and occasionally mentioned in this paper.
 A detailed review of these topics is well outside the scope of this paper, however.
 Some of the model-approaches not based on vector magnetograms are occasionally
 used to test the nonlinear force-free extrapolation codes described here.
 \subsection{Why do we need nonlinear force-free fields?}
\begin{figure}
  \includegraphics[width=10cm,height=14cm]{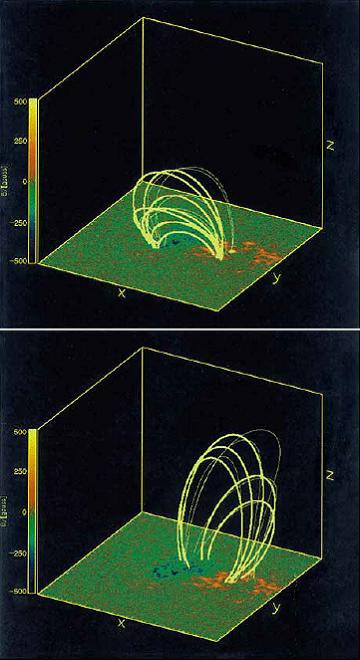}
 \caption{Linear force-free field model for NOAA 7986 with the best fit for $\alpha$.
  The upper panel shows a group of loops with $\alpha=2.5$ and the lower
  panel another group of loops with $\alpha=-2.0$. The different optimal
  values of the linear force-free parameter within one active region are
  a contradiction to the linear assumption ($\alpha$ constant) and tell us
  that a consistent modelling of this active regions requires a nonlinear
  force-free approach.
 [This figure has been originally published as Fig. 7 in \cite{wiegelmann:etal02}.
 Used with permission of Springer.]}
  \label{figure1}
\end{figure}

\begin{figure}
  \includegraphics[width=15cm,height=15cm]{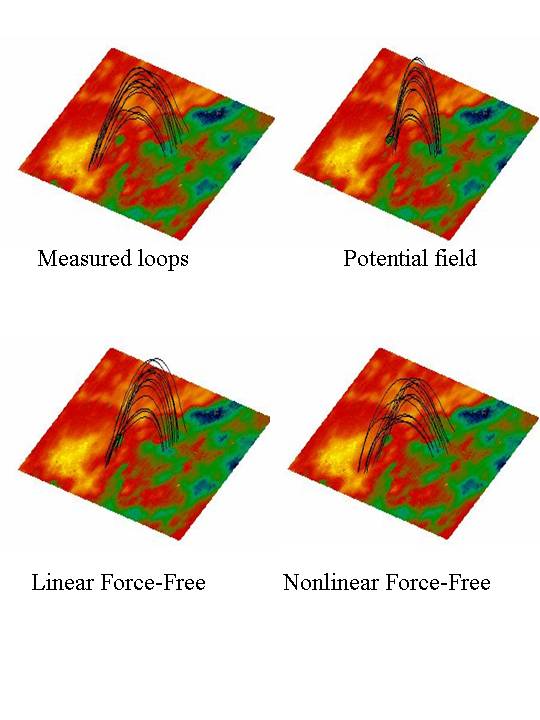}
 \caption{Magnetic field structure of the newly developed active region NOAA 9451.
  Direct measurements of the field have been compared with a potential, linear  and
  nonlinear force-free model. Best agreement has been found for the nonlinear model.
 [This figure has been originally published as part of Fig. 1 in
 \cite{wiegelmann:etal05}. Used with permission of A \& A.]}
  \label{figure2}
\end{figure}

 \begin{itemize}
 \item A comparison of global potential magnetic field models with TRACE-images
 by \cite{schrijver:etal05a} revealed that significant nonpotentially occurs
 regulary in active regions, in particular when new flux has emerged in or
 close to the regions.
 \item Usually $\alpha$ changes in space, even inside one active region.
 This can be seen, if we try to fit for the optimal linear force-free
 parameter $\alpha$ by comparing field lines with coronal plasma structures.
 An example can be seen in \cite{wiegelmann:etal02} where stereoscopic
 reconstructed loops by \cite{aschwanden:etal99} have been compared
 with a linear force-free field model. The optimal value of $\alpha$ changes
 even sign within the investigated active regions, which is a contradiction to
 the $\alpha={\rm constant}$ linear force-free approach
 (see Fig. \ref{figure1}).
 \item Photospheric $\alpha$ distributions derived from vector magnetic
 field measurements by Eq. (\ref{alpha0direct}) show as well
 that $\alpha$ usually changes within an active region
 \citep[see, e.g.][]{regnier:etal02}.
 \item Potential and linear force-free fields are too simple
 to estimate the free magnetic energy and magnetic topology accurately.
 The magnetic energy of linear force-free fields is unbounded in a
 halfspace \citep{seehafer78} which makes this approach unsuitable for
 energy approximations of the coronal magnetic field. Potential fields
 have a minimum energy for an observed line-of-sight photospheric magnetic
 field. An estimate of the excess of energy a configuration has above that
 of a potential field is an important quantity which might help to
 understand the onset of flares and coronal mass ejections.
 \item A direct comparison of measured fields in a newly developed
 active region by \cite{solanki:etal03} with extrapolations from the
 photosphere with a potential, linear and nonlinear force-free model
 by \cite{wiegelmann:etal05}
 showed that nonlinear fields are more accurate than simpler models.
 Fig. \ref{figure2} shows some selected magnetic field lines for
 the original measured field and extrapolations from the photosphere
 with the help of a potential, linear and nonlinear force-free model.
 \end{itemize}
 These points tell us that nonlinear force-free modelling is required
 for an accurate reconstruction of the coronal magnetic field. Simpler
 models have been used frequently in the past.
 Global potential fields provide some information of the coronal
 magnetic field structure already, e.g. the location of coronal
 holes. The generic case of force-free coronal magnetic field models are
 nonlinear force-free fields, however. Under generic we understand that
 $\alpha$ can (and usually will) change in space, but this approach
 also includes the special cases $\alpha={\rm constant}$ and $ \alpha=0$.
 Some active regions just happen to be more potential (or linear force-free)
 and if this is the case they can be described with simpler models.
 Linear force-free models might provide a rough estimate of
 the true 3D magnetic field structure if the nonlinearity is weak.
 The use of simpler models was often justified due to limited observational
 data, in particular if only the line-of-sight photospheric magnetic field
 has been measured.

 While the assumption of nonlinear force-free fields is well accepted
 for the coronal magnetic fields in active regions, this is not true
 for the photosphere. The photospheric plasma is a finite $\beta$
 plasma and nonmagnetic forces like pressure gradient and gravity
 cannot be neglected here. As a result electric currents have a component
 perpendicular to the magnetic field, which contradicts the force-free
 assumption. We will discuss later, how these difficulties can be
 overcome.

\section{Nonlinear force-free codes}
Different methods have been proposed to extrapolate nonlinear force-free
fields from photospheric vector magnetic field measurements.
\begin{enumerate}
\item The Grad-Rubin method was proposed for fusion plasmas by \cite{grad:etal58}
and first applied to coronal magnetic fields by \cite{sakurai81}.
\item The upward integration method was proposed by \cite{nakagawa74}
 and encoded by \cite{wu:etal85}.
\item The MHD relaxation method was proposed for general MHD-equilibria by
\cite{chodura:etal81} and applied to force-free coronal magnetic fields by
\cite{mikic:etal94}.
\item The optimization approach was developed by \cite{wheatland:etal00}.
\item The boundary element (or Greens function like) method was developed by
\cite{yan:etal00}.
\end{enumerate}
\subsection{Grad-Rubin method}
 The Grad-Rubin method reformulates the nonlinear force-free equations in
 such a way, that one has to solve a well posed boundary value problem.
 This makes this approach also interesting for a mathematical investigation
 of the structure of the nonlinear force-free equations.
 \cite{bineau72} demonstrated that the used boundary conditions (vertical
 magnetic field on the photosphere and $\alpha$ distribution at one polarity)
 ensure, at least for small values of $\alpha$ and weak nonlinearities the
 existence of a unique nonlinear force-free solution. A detailed analysis of
 the mathematical problem of existence and uniqueness of nonlinear force-free
 fields is outside the scope of this review and can be found e.g. in
 \cite{amari:etal97,amari:etal06}.

 The method first computes a potential field, which can be obtained from
 the observed line-of-sight photospheric magnetic field (say $B_z$ in
 Cartesian geometry) by different methods, e.g., a Greens function method as
 described in \cite{aly89}. It is also popular to use linear force-free solvers,
 e.g. as
 implemented by \cite{seehafer78,alissandrakis81} with the linear force-free
 parameter $\alpha=0$ to compute the initial potential field. The transverse
 component of the measured magnetic field is then used to compute the
 distribution of $\alpha$ on the photosphere by Eq. (\ref{alpha0direct}).
 While $\alpha$ is described this way on the entire photosphere, for both polarities,
 a well posed boundary value problem requires that the $\alpha$ distribution becomes only
 described for one polarity. The basic idea is to
 iteratively calculate $\alpha$ for a given ${\bf B}$ field from
 (\ref{alphaeq}), then calculate the current via (\ref{jparb})
 and finally update ${\bf B}$ from the Biot-Savart problem
 (\ref{amperealpha}).
 These processes are repeated until the full current as prescribed by
 the $\alpha$-distribution has been injected into the magnetic field and the updated
 magnetic field configuration becomes stationary in the sense that eventually the
 recalculation of the magnetic field with Amperes law does not change the configuration
 anymore. To our knowledge the  Grad-Rubin approach has been
 first implemented by \cite{sakurai81}. $\alpha$ has been prescribed
 on several nodal points along a number of magnetic field lines of
 the initial potential field. The method used a finite-element-like
 discretization of current tubes associated with magnetic field lines.
 Each current tube was divided into elementary current tubes of
 cylindrical shape. The magnetic field is updated with Ampere's law
 using a superposition of the elementary current tubes. The method was
 limited by the number of current-carrying field lines, nodal-points and
 the corresponding number of nonlinear equations $(N^9)$ to solve
 with the available computer resources more than a quarter century ago.

 Computer resources have increased rapidly since the first NLFFF-implementation by
 \cite{sakurai81} and about a decade ago the Grad-Rubin method
 has been implemented on a finite difference grid
 by \cite{amari:etal97,amari:etal99}.
 This approach decomposes the equations (\ref{forcebal}-\ref{solenoidal})
 into a hyperbolic part for evolving $\alpha$ along
 the magnetic field lines and an elliptic one to iterate the updated magnetic
 field from Amperes law. For every iteration step $k$ one has to solve iteratively
 for:

\begin{eqnarray}
 {\bf B}^{(k)} \cdot \nabla \alpha^{(k)} &= &0 \\
 \alpha^{(k)}|_{S^{\pm}} &=& \alpha_{0\pm} \\
 \end{eqnarray}

 which evolves $\alpha$ in the volume and

 \begin{eqnarray}
 \nabla \times {\bf B}^{(k+1)} &=& \alpha^{(k)} {\bf B}^{(k)} \\
 \nabla \cdot {\bf B}^{(k+1)} &=& 0 \\
 B^{(k+1)}_z|_{S^{\pm}}  &=& B_{z0} \\
 {\rm \lim_{|r|\rightarrow\infty}}|{\bf B}^{(k+1)}| &=&0,
 \end{eqnarray}

 where $\alpha_{0\pm}$ corresponds to the photospheric distribution of
 $\alpha$ for either on the positive or the negative polarity.
 The Grad-Rubin method as described in \cite{amari:etal97,amari:etal99}
  has been applied to investigate particular active regions in \cite{bleybel:etal02}
  and a comparison of the extrapolated field with
 2D projections of plasma structures as seen in H$\alpha$, EUV and X-ray has been done in
\cite{regnier:etal02,regnier:etal04}. The code has also been used to investigate
mutual and self helicity in active regions by \cite{regnier:etal05} and to flaring
active regions by \cite{regnier:etal06}.

 A similar approach as done by \cite{sakurai81} has been implemented by \cite{wheatland04}.
 The implemented method computes the magnetic field directly on the numerical grid
 from Ampere's law. This is somewhat simpler and faster as Sakurai's approach which
 required solving a large system of nonlinear equations for this aim. The implementation
 by \cite{wheatland04} has, in particular, been developed with the aim of parallelization.
 The parallelization approach seems to be effective due to a limited number of inter-process
 communications. This is possible because as the result of the linearity
 of Ampere's law the contributions
 of the different current carrying field lines are basically independent from each other.
 In the original paper Wheatland reported problems for large currents on the field lines.
 These problems have been related to an error in current representation of the code and
 the corrected code worked significantly better, see also \cite{schrijver:etal06}.
 The method has been further developed in \cite{wheatland06}. This newest
 Wheatland-implementation
 scales with the number of grid points $N^4$ for a $N^3$ volume, rather than $N^6$ for
 the earlier \cite{wheatland04} implementation. The main new development is a faster
 implementation of the current-field iteration. To do so the magnetic field has been separated
 into a current-free and a current carrying part at each iteration step. Both parts are
 solved using a discrete Fast Fourier Transformation, which imposes the
 required boundary conditions implicitly. The code has been parallelized on
 shared memory distributions with OpenMP.

 \cite{amari:etal06} developed two new versions of their Grad-Rubin code.
 The first version is a finite difference method and the code was called
 'XTRAPOL'. This code prescribe the coronal magnetic field with the
 help of a vector potential ${\bf A}$.
 The code has obviously it's heritage from the earlier implementation
 of \cite{amari:etal99}, but with several remarkable differences:
 \begin{itemize}
 \item The code includes a divergence cleaning routine, which takes care
 about $\nabla \cdot {\bf A}=0$. The condition
 $\nabla \cdot {\bf A}=0$ is fulfilled with
 high accuracy in the new code $10^{-9}$ compared to $10^{-2}$
 in the earlier implementation.
 \item The lateral and top boundaries are more flexible compared to the earlier
 implementation and allow a finite  $B_n$ and non zero
 $\alpha$-values for one polarity on all boundaries. This treats the whole boundary
 (all six faces) as a whole.
 \item The slow current input as reported for the earlier implementation,
 which lead to a two level iteration, has
 been replaced. Now the whole current is injected at once and only the inner
 iteration loop of the earlier code remained in the new version.
 \item The computation of the $\alpha$ characteristics has been improved with
 an adaptive Adams-Bashforth integration scheme \citep[see][for details]{press02}.
 \item The fixed number of iteration loops have been replaced by a
 quantitative convergence criterium.
\end{itemize}
 In the same paper \cite{amari:etal06} introduced another Grad-Rubin approach
 based on finite elements, which they called 'FEMQ'. Different from
 alternative implementation this code does not use a vector potential
 but iterates the coupled divergence and curl system, which is
 solved with the help of a finite element discretization.
 The method transforms the nonlinear force-free equations into a global
 linear algebraic system.

 \cite{inhester:etal06} implemented a Grad-Rubin code on a
 finite element grid with staggered field components
 (see \cite{Yee:1966}) which uses discrete Whitney forms
 (\cite{Bossavit:1988}).
 Whitney forms allow to transform standard vector analysis
 (as the differential operators gradient, curl and divergence)
 consistently into the discrete space used for numerical computations.
 Whitney forms contain four types of finite elements (form 0-3).
 They can be considered as a discrete approximation of differential
 forms. The finite element base may consist of polynominals of
 any order. In its simplest form, the
 0-forms have as parameters the function values at the vertices of the
 cells and are linearly interpolated within each cell. 1-forms are a
 discrete representation of a vector field defined on the cell edges.
 2-forms are defined as the field component normal to the surfaces of
 the cells. The 3-forms are finite volume elements for a scalar
 function approximation, which represents the average of a scalar
 over the entire cell. The four forms are related to each other
 by GRAD (0 to 1 form), CURL (1 to 2 form) and DIV (2 to 3 form).
 As for continous differential forms, double differentiation
 (CURL of GRAD, DIV of CURL) give exactly zero, independent of the
 numerical precision. A dual grid, shifted by half a grid size
 in each axis, was introduced in order to allow for
 Laplacians. Whitney forms on the dual grid are related to forms
 on the primary grid in a consistent way.

 The Grad-Rubin implementation uses a vector potential representation of
 the magnetic field, where the vector potential is updated with a
 Poisson equation in each iteration step. The Poisson equation is
 effectively solved with the help of a multigrid solver. The main computing
 time is spend to distribute $\alpha$ along the field lines with
 (\ref{alphaeq}). This seems to be a general property of Grad-Rubin
 implementations. One can estimate the scaling of
 (\ref{alphaeq}) by $\propto N^4$, where the number of field lines to
 compute is $\propto N^3$ and the length of a field line $\propto N$.
 The Biot-Savart step (\ref{amperealpha}) solved with FFT or multigrid
 methods scales only with $\propto N \log N$. Empirical tests show that
 the number of iteration steps until a stationary state is reached does not
 depend on the number of grid points $N$ for Grad-Rubin solvers.
 We have explained before, that the Grad-Rubin implementation
 requires the prescription of $\alpha$ only for one polarity to have a well
 posed mathematical problem. The \cite{inhester:etal06} implementation allows
 these choice of boundary conditions as a special case. In general one does
 not need to make the distinction between  $(\partial V)^+$ and
  $(\partial V)^-$ in the new implementation. A well posed mathematical problem
  is still ensured, however, in the following way. Each boundary value
  of $\alpha$ is attached with a weight. The final version of $\alpha$
 on each field line is then determined by a weighted average of the
 $\alpha$-values on both endpoints of a field lines. By this way the
 influence of uncertain boundary values, e.g. on the side walls and
 imprecise photospheric measurements  can be suppressed.

\subsection{Upward integration method}
 The basic equations for the upward integration method
 (or progressive extension method) have been published already
 by \cite{nakagawa74} and a corresponding code has been developed by
 \cite{wu:etal85,wu:etal90}. The upward integration method is a straight forward
 approach to use the nonlinear force-free equations directly to extrapolate the
 photospheric magnetic field into the corona.
 To do so one
 reformulates the force-free equations (\ref{forcebal}-\ref{solenoidal})
 in order to extrapolate the measured
 photospheric magnetic field vector into the solar corona.

As a first step the magnetic field vector on the lower boundary
${\bf B}_0(x,y,0)$ is used to compute the $z$-component of the
electric current $\mu_0 j_{z0}$
with Eq. (\ref{j_photo}) and the photospheric $\alpha$-distribution
(say $\alpha_0$) by Eq. (\ref{alpha0direct}).
With the help of Eq. (\ref{jparb}) we
calculate the $x$ and $y$-component of the current density

\begin{eqnarray}
\mu_0 j_{x0 } & = & \alpha_0\, B_{x0},\\
\mu_0 j_{y0}  & = & \alpha_0\, B_{y0}.
\end{eqnarray}

We now use Eq. (\ref{solenoidal}) and the $x$ and $y$-component
of Eq. (\ref{ampere}) to obtain expressions for the $z$-derivatives of
all three magnetic field components in the form

\begin{eqnarray}
\frac{\partial B_{x0}}{\partial z} & = & \mu_0 j_{y0}+\frac{\partial B_{z0}}{\partial x}, \\
\frac{\partial B_{y0}}{\partial z} & = &\frac{\partial B_{z0}}{\partial y}- \mu_0 j_{x0},\\
\frac{\partial B_{z0}}{\partial z} & = & -\frac{\partial B_{x0}}{\partial x}-
\frac{\partial B_{y0}}{\partial y}.
\end{eqnarray}

The idea is to integrate this set of equations numerically upwards in $z$ by
repeating the previous steps at each height. As a result we get in principle the 3D
magnetic field vector in the corona. While this approach is straight forward, easy
to implement and computational fast (no iteration is required), a serious drawback
is that it is unstable. Several authors
\citep[e.g.,][]{cuperman:etal90,amari:etal97} pointed out that the formulation of
the force-free equations in this way is unstable because it is based on an ill-posed
mathematical problem. In particular one finds that exponential growth of the
magnetic field with increasing height is a typical behaviour.
%
What makes this  boundary value problem ill-posed is that the solution
does not depend continuously on the boundary data. Small changes or inaccuracies
in the measured boundary data lead to a divergent extrapolated field
\citep[see][for a more detailed discussion]{low:etal90}.
As pointed out by Low and Lou meaningful boundary conditions are required also
on the outer boundaries of the computational domain. It is also possible
to prescribe open boundaries in the sense that the magnetic field vanishes at
infinity. This causes an additional problem for the upward integration method,
because the method transports information only from the photosphere upwards and
does not incorporate boundary information on other boundaries or at infinity.
Attempts have been made to regularize the
method \citep[e.g.,][]{cuperman:etal91,demoulin:etal92}, but cannot be considered as
fully successful.

 \cite{wu:etal90a} compared the Grad-Rubin method in the implementation
 of \cite{sakurai81} with the upward integration method in the implementation
 of \cite{wu:etal90} \footnote{The authors used a somewhat different nomenclature:
 The upward integration method was called 'progressive extension method' and
 the Grad-Rubin method 'iterative method'. That time the term 'iterative method'
 was reasonable because Grad-Rubin was the only iterative approach available,
 but now, 17 years later, several other iterative methods are available to
 compute nonlinear force-free fields.}. The comparison showed qualitatively
 similar results for extrapolations from an observed magnetogram, but quantitatively
 differences. The NLFFF-computations have been very similar to potential field
 extrapolations, however, too. One reason for this behaviour was, that the method
 of \cite{sakurai81} is limited to small values of $\alpha$ and an 'by eye'
 comparison shows that the corresponding NLFFF field is very close
 to a potential field configuration.
 The field computed with the upward integration method deteriorated
 if the height of the extrapolation exceeded a typical horizontal
 scale length.

 The upward integration method has been recently reexamined by \cite{song:etal06} who
 developed a new formulation of this approach. The new implementation
 uses smooth continuous functions and the equations are solved in asymptotic
 manner iteratively. The original upward integration equations are reformulated into
 a set of ordinary differential equations and uniqueness of the solution
 seems to be guarantied at least locally. While \cite{demoulin:etal92}
 stated that 'no further improvement has been obtained with other types of
 smoothing functions' the authors of \cite{song:etal06} point out that
 the transformation of the original partial differential equations into
 ordinary ones eliminates the growing modes in the upward
 integration method, which have been reported before in \cite{wu:etal90} and subsequent
 papers. The problem that all three components of the photospheric magnetic field and the
 photospheric $\alpha$ distribution has to be prescribed in a consistent way remains
 in principle, but some compatibility conditions to compute a slowly varying
 $\alpha$ have been provided by \cite{song:etal06}. These compatibility conditions
 are slightly different for real photospheric observations and tests with smooth
 boundaries extracted from semi-analytic equilibria.
 For the latter kind of problems the new formulation provided reasonable results
 with the standard test equilibrium  found by \cite{low:etal90}. The method seems
 to be also reasonable fast. Of course further tests with more sophisticated
 equilibria and real data are necessary to evaluate this approach in more detail.
\subsection{MHD relaxation}
 MHD relaxation codes \citep[e.g.,][]{chodura:etal81} can be
 applied to solve nonlinear force-free fields as well.
 The idea is to start with a suitable
magnetic field which is not in equilibrium and to relax it into a force-free state.
This is done by using the MHD equations in the following form:

\begin{eqnarray}
\nu  {\bf v }&=& (\nabla \times {\bf B}) \times {\bf B} \label{eqmotion}\\
{\bf E} + {\bf v} \times {\bf B} &=& {\bf 0 } \label{iohms}\\
\frac{\partial {\bf B}}{\partial t} &=&-\nabla \times {\bf E} \label{faraday} \\
\nabla \cdot {\bf B} &=& 0 \label{sole2},
\end{eqnarray}

where $\nu$ is a viscosity and ${\bf E}$ the electric field. As the MHD-relaxation
aims for a quasi physical temporal evolution of the magnetic field from a
non-equilibrium towards a (nonlinear force-free) equilibrium this method is also
called 'evolutionary method' or 'magneto-frictional method'.
 The basic idea is that the velocity field in the
equation of motion (\ref{eqmotion}) is reduced during the relaxation process. Ideal
Ohm's law (\ref{iohms}) ensures that the magnetic connectivity remains unchanged
during the relaxation. The artificial viscosity $\nu$ plays the role of a relaxation
coefficient which can be chosen in such way that it accelerates the approach to the
equilibrium state. A typical choice is

\begin{equation}
\nu =  \frac{1}{\mu} \, |{\bf  B}|^2
\label{nudef}
\end{equation}

with
$\mu=$ constant. Combining Eqs. (\ref{eqmotion}), (\ref{iohms}),
(\ref{faraday}) and (\ref{nudef}) we get an equation for the
evolution of the magnetic field during the
relaxation process:

\begin{equation}
\frac{\partial {\bf B}}{\partial t} =\mu \; {\bf F_{\rm MHD}}
\label{relaxinduct}
\end{equation}

with

\begin{equation}
{\bf F_{\rm MHD}} =\nabla \times
\left(\frac{\left[(\nabla \times {\bf B}) \times {\bf B }\right]
\times {\bf B}}{B^2}  \right).
\label{def_fmhd}
\end{equation}

This equation is then solved numerically starting with a given initial condition for
${\bf B}$, usually a potential field. Equation (\ref{relaxinduct}) ensures that Eq.
(\ref{sole2}) is satisfied during the relaxation if the initial magnetic field
satisfies it. \footnote{As we will see below the 'optimization' approach leads to a
similar iteration equations for the magnetic field, but a different artificial
driving force ${\bf F}$.} The difficulty with this method is that it cannot be
guaranteed that for given boundary conditions and initial magnetic field (i.e. given
connectivity), a smooth force-free equilibrium exists to which the system can relax.
If such a smooth equilibrium does not exist the formation of current sheets is to be
expected which will lead to numerical difficulties. Therefore, care has to be taken
when choosing an initial magnetic field.

 \cite{yang:etal86} developed a magneto frictional method which represent the
 magnetic field with the help of Euler (or Clebsch) potentials.

\begin{equation}
 {\bf B}=\nabla g \times \nabla h,
\end{equation}

 where the potentials $g$ and $h$ are scalar functions. The general
 method has been developed for three dimensional fields and iterative
 equations for $g(x,y,z)$ and $h(x,y,z)$ have been derived. The Clebsch representation
 automatically ensures $\nabla \cdot {\bf B}=0$. The method has been
 explicitly tested in the paper by \cite{yang:etal86} with the help of an
 equilibrium with one invariant coordinate. In principle it should be possible to use
 this representation for the extrapolation of nonlinear force-free fields,
 but we are not aware of a corresponding implementation. Due to the discussion
 in \cite{yang:etal86} a difficulty seems to be that one needs to specify boundary
 conditions for the potentials, rather than for the magnetic fields. It seems
 in particular to be difficult to find boundaries conditions for potentials which
 correspond to the transverse component of the photospheric magnetic field vector.
 One problem is that boundary conditions for $g$ and $h$ prescribe the
 connectivity. Every field line can be labelled by its $(g,h)$ values.
 Hence boundary values for $g$ and $h$ establish foot point relations
 although the field is not known yet.

 The MHD-relaxation (or evolutionary) method has been implemented by
 \cite{mikic:etal94,mcclymont:etal97} based on the time dependent MHD-code
 by \cite{mikic:etal88}. The code uses a nonuniform mesh and the region of
 interested is embedded in a large computational domain to reduce the influence
 of the lateral boundaries. The method has been applied to extrapolate the
 magnetic field above an active region by \cite{jiao:etal97}.
 The computations have been carried out with a resolution of
  the order of $100^3$ points. A super-computer was required for these
  computations that time (10 years ago), but due to the rapid increase
  of computer speed and memory within the last decade this restriction is very
 probably not valid anymore.

 \cite{roumeliotis96} developed the so called stress and relax method.
 In this approach the initial potential field becomes disturbed by
 the observed transverse field component on the photosphere. The boundary
 conditions are replaced in subsequently in several small steps and always
 relaxed with a similar MHD-relaxation scheme as described above towards
 a force-free equilibrium. The code by \cite{roumeliotis96}
 has implemented a function $w(x,y)$ which allows to give a lower weight to
 regions where the transverse photospheric field has been measured with
 lower accuracy. Additional to the iterative equations as discussed above,
 the method includes a resistivity $\eta$ (or diffusivity) by adding
 a term $\eta {\bf j}$ on the right hand site of Ohms law (\ref{iohms}).
 This relaxes somewhat
 the topological constrains of ideal MHD relaxation, because a finite
 resistivity allows a kind of artificial reconnection and corresponding
 changes of the initial potential field topology.
 The method has been tested with a force-free equilibrium found by \cite{klimchuk:etal92}
 and applied to an active region measured with the MSFC vector-magnetograph.

 The stress and relax method has been revisited by \cite{valori:etal05}.
 Different from the earlier implementation by \cite{roumeliotis96}
 the new implementation uses directly the magnetic field, rather than
 the vector potential in order to keep errors from taking numerical
 deviations from noisy magnetograms minimal.
 The solenoidal condition is controlled by a diffusive approach
 by \cite{dedner:etal02} which
 removes effectively a numerically created finite divergence of the
 relaxed magnetic field. The new implementation
 uses a single stress step, rather than the multiple small stress used
 by \cite{roumeliotis96} to speed up the computation.

 The single step stress and relax method is connected with a suitable control
 of artificial plasma flows by the Courant criterium. The authors reported that
 a multi-step and single-step implementation do not reveal significant differences.
 The numerical implementation is based on the time-depended full MHD-code
 'AMRVAC' by \cite{keppens:etal03}.
 \cite{valori:etal05} tested their nonlinear force-free implementation
 with a numerically constructed nonlinear force-free twisted loop computed
 by \cite{toeroek:etal03}.

\subsection{Optimization approach}
The optimization approach has been developed in \citet{wheatland:etal00}. The
solution is found by minimizing the functional

\begin{equation}
L=\int_{V} \left[B^{-2} \, |(\nabla \times {\bf B}) \times {\bf B}|^2 +|\nabla \cdot
{\bf B}|^2\right] \; d^3V.
\label{defL}
\end{equation}

 Obviously, $L$ is bound from below by $0$. This bound is attained if the magnetic
field satisfies the force-free equations
 (\ref{forcebal})-(\ref{solenoidal}).

 By taking the functional derivatives with respect to some iteration
 parameter $t$ we get:

\begin{equation}
\Rightarrow \frac{1}{2} \; \frac{d L}{d t}=-\int_{V} \frac{\partial {\bf
B}}{\partial t} \cdot {\bf \tilde{F}} \; d^3x -\int_{S} \frac{\partial {\bf
B}}{\partial t} \cdot {\bf \tilde{G}} \; d^2x \label{minimize1}
\end{equation}

 with

\begin{eqnarray}
{\bf F} & = & \nabla \times \left(\frac{\left[(\nabla \times \bf B) \times \bf B
\right]
\times \bf B}{B^2}  \right) \nonumber\\
& &  + \;  \left\{ -\nabla \times \left(\frac{((\nabla \cdot \bf B) \; \bf B)
\times \bf B}{B^2} \right) \right.  \nonumber\\
& & - \bf \Omega \times (\nabla \times \bf B) -\nabla(\bf \Omega \cdot \bf B) \nonumber\\
& & + \left. \bf \Omega(\nabla \cdot \bf B) + \Omega^2 \; \bf B \right\}
\label{wheateq}
\end{eqnarray}

\begin{equation}
\bf \Omega=B^{-2} \;\left[(\nabla \times \bf B) \times \bf B-(\nabla \cdot \bf B) \;
\bf B \right]
\end{equation}

The surface term vanishes if the magnetic field vector is kept constant on the
surface, e.g. prescribed from photospheric measurements. In this case $L$ decreases
 monotonically if the magnetic field is iterated by

\begin{equation}
\frac{\partial \bf B}{\partial t}  = \mu \; {\bf F}.
\end{equation}

Let us remark that ${\bf F_{\rm MHS}}$ as defined in Equation (\ref{def_fmhd}) and
used for MHD-relaxation is identical with the first term on the right-hand-side of
Eq. (\ref{wheateq}), but Eq. (\ref{wheateq}) contains additional terms.

For this method the vector field ${\bf B}$ is not necessarily
so\-le\-noi\-dal during the computation, but will be divergence-free
if the optimal state with $L=0$ is reached. A disadvantage of the method
is that it cannot be guaranteed that this optimal state is indeed
reached for a given initial field and boundary conditions. If this is
not the case then the resulting ${\bf B}$ will either be not force-free
or not solenoidal or both.

 McTiernan has implemented the optimization approach basically as described in
 \cite{wheatland:etal00} in IDL (see \cite{schrijver:etal06} for a brief
 description of the McTiernan implementation.) This code allows the use of
 a non-uniform computational grid. In a code inter-comparison by
 \cite{schrijver:etal06} the IDL optimization code by McTiernan was about a
 factor of $50$ slower compared to an implementation in parallelized C by
 \cite{wiegelmann04}.  To our knowledge McTiernan has translated his IDL-code
 into FORTRAN in the meantime for faster computation (personal communication
 on the NLFFF-workshop Palo-Alto, june 2006. See also \cite{metcalf:etal07}.)

 Several tests have been performed with the optimization
 approach in \cite{wiegelmann:etal03}.
 It has been investigated how the unknown lateral
 and top boundary influence the solution. The original optimization approach by
 \cite{wheatland:etal00} has been extended towards more flexible boundary-conditions, which
 allow $\frac{\partial {\bf B}}{\partial t} \not=0$ on the lateral and top boundaries. This has
 been made with the help of the surface integral term in (\ref{minimize1}) and led to
 an additional term $\frac{\partial {\bf B}}{\partial t} = \mu {\bf G}$ on the boundaries.
 This approached improved the performance
 of the code for cases, where only the bottom boundary was prescribed. No improvement
 was found for a slow multi-step replacement of the boundary and this possibility has been
 abandoned in favour of a single step method.

 It has been also investigated how noise influences the optimization code
 and this study revealed that noise in the vector magnetograms leads to  less accurate
 nonlinear force-free fields.

\cite{wiegelmann04} has reformulated the optimization principle by introducing
weighting functions
One defines the functional

\begin{equation}
L=\int_{V} \;  \left[w \; B^{-2} \, |(\nabla \times {\bf B}) \times {\bf B}|^2 +w \;
|\nabla \cdot {\bf B}|^2\right] \; d^3x \label{defL1},
\end{equation}

where $w(x,y,z)$ is a weighting function. It is obvious that (for $w>0$) the
force-free equations (\ref{forcebal}-\ref{solenoidal}) are fulfilled when L is equal
zero. Minimization of the functional (\ref{defL1}) lead to:

\begin{equation}
\frac{\partial {\bf B}}{\partial t} =\mu {\bf \tilde{F}}
\label{iterateB},
\end{equation}
\begin{eqnarray}
{\bf \tilde{F}}&=& w \; {\bf F} +({\bf \Omega_a} \times {\bf B} )\times \nabla w
+({\bf \Omega_b} \cdot {\bf B}) \; \nabla w \\
 {\bf \Omega_a} &=& B^{-2} \;\left[(\nabla \times \bf B) \times {\bf B} \right] \\
 {\bf \Omega_b} &=& B^{-2} \;\left[(\nabla \cdot \bf B) {\bf B} \right],
\end{eqnarray}

with ${\bf F}$ as defined in (\ref{wheateq}).
With $w(x,y,z)=1$ this approach reduces to the  \cite{wheatland:etal00} method
 as described above.
The weighting function is useful if only the bottom boundary data are known. In this
case we a buffer boundary of several grid points towards the lateral and
top boundary of the computational box is introduced. The weighting function is chosen constant in
the inner, physical domain and drop to $0$ with a cosine profile in the buffer
boundary towards the lateral and top boundary of the computational box.
 In \cite{schrijver:etal06} some tests have been made with different
 weighting functions for the force-free and solenoidal part of the functional
 $(\ref{defL1})$, but the best results have been obtained if both terms
 got the same weight. The computational implementation involves the
 following steps.
\begin{enumerate}
\item Compute start equilibrium (e.g. a potential field) in the computational box.
\item Replace the bottom boundary  with the vector magnetogramm.
\item Minimize the
functional (\ref{defL1}) with the help of Eq. (\ref{iterateB}). The continuous form
of (\ref{iterateB}) guaranties a monotonically decreasing $L$. This is as well
ensured  in the discretized form  if the iteration step $dt$ is sufficiently small.
The code checks if $L(t+dt) < L(t)$ after each time step. If the condition is not
fulfilled, the iteration step is repeated with $dt$ reduced by a factor of 2. After
each successful iteration step we increase $dt$ slowly by a factor of $1.01$ to
allow the time step to become as large as possible with respect to the stability
condition. \item The iteration stops if $L$ becomes stationary. Stationarity is
assumed if $\frac{\partial L}{\partial t}/L < 1.0 \cdot 10^{-4}$ for $100$
consecutive iteration steps.
\end{enumerate}
The program has been tested with the semi-analytic nonlinear force-free
configuration by \cite{low:etal90} and \cite{titov:etal99} in
\cite{wiegelmann:etal06a}. The code has been applied to extrapolate the coronal
magnetic field in active regions in \cite{wiegelmann:etal05,wiegelmann:etal05b}.

 A finite element optimization approach has been implemented by
 \cite{inhester:etal06} using the Whitney elements as for the Grad-Rubin
 code (which has been described above). The optimization method uses
 exactly the same staggered finite element grid as described above,
 which is different from the finite difference grids used in the
 earlier implementations by
 \cite{wheatland:etal00,wiegelmann:etal03,wiegelmann04}. Another difference
 is that earlier implementations discretized the analytical derivative
 of the functional $L$ (\ref{wheateq}), while
 the new code takes the numerical more consistent derivative of
 the discretized function $L$. All other  implementations used a simple
 Landweber scheme for updating the magnetic field, which is replaced
 here by an unpreconditioned conjugate gradient iteration, which at every
 time step performs an exact line search to the minimum of $L$ in the
 current search direction and additional selects an improve search
 direction instead of the gradient of the functional $L$. To do so
 the Hessian matrix of the functional $L$ is computed during every
 iteration step. An effective computation of the Hessian matrix is
 possible, because the reformulated function $L(s)$ is a fourth order
 polynomial in ${\bf B}$ and all five polynomial coefficients can
 be computed in one go. The code has been tested with \cite{low:etal90}
 and the result of twisted loop computations of the Grad-Rubin
 implementation on the same grid.

The optimization code in the
implementation of \cite{wiegelmann04} has recently be extended towards using a
multi-scale implementation. The main difference from the original code are (see also
\cite{metcalf:etal07}):

\begin{itemize}
 \item The method is not full multigrid, but computes the solution on different grids
 only once, e.g., something like $ 50^3, \;  100^3, \; 200^3$.
 \item The main idea is to get a better (than potential field) start equilibrium
on the full resolution box.
 \item Solution of smaller grids are interpolated onto larger grids as
initial state for the magnetic field in the computational domain of the next larger
box.
\end{itemize}
The multiscale implementation has been tested as part of a code-inter-comparison
test in \cite{metcalf:etal07} with the help of solar-like reference model computed
by \cite{vanballegooijen04,vanballegooijen:etal07}.

 The optimization approach has recently be implemented  in
 spherical geometry by \cite{wiegelmann07} and tested with
 \cite{low:etal90}. The original longitudinal symmetric Low
 and Lou solution has been shifted by $1/4$ of a solar radius to
 test the code without any symmetry with respect to the Suns surface.
 The numerical implementation is very similar as the Cartesian implementation
 described in \cite{wiegelmann04}. The spherical implementation
 converged fast for low latitude regions, but the computing time
 increased significantly if polar regions have been included. It has
 been suggested to implement the code on a so called 'Yin and Yang'
 grid as developed by \cite{kageyama:etal04} to reduce the computing time.
 The 'Yin and Yang' grid is suitable for massive parallelization,
 which is necessary for full sphere high resolution NLFFF-computations.
\subsection{Boundary element or Greens function like method}
 The boundary integral method has been developed by
 \cite{yan:etal00}. The method relates the measured
 boundary values with the nonlinear force-free field
 in the entire volume by:

 \begin{equation}
 c_i {\bf B_i}= \oint_S \left({\bf {\bar Y}} \frac{\partial {\bf B}}{\partial n}
 - \frac{\partial {\bf {\bar Y}}}{\partial n} {\bf B_0} \right) \; {\bf dS},
 \label{yan1}
 \end{equation}

 where $c_i=1$ for points in the volume and  $c_i=1/2$ for
 boundary points and ${\bf B_0}$ is the measured vector
 magnetic field on the photosphere.
 The auxiliary vector function is defined as

 \begin{equation}
 {\bf \bar{Y}}={\rm diag} \left(\frac{cos(\lambda_x r)}{4 \pi r},\frac{cos(\lambda_y r)}{4 \pi r},
 \frac{cos(\lambda_z r)}{4 \pi r}  \right)
\end{equation}

  and the $\lambda_i, \, (i=x,y,z)$ are computed in the original approach by
 \cite{yan:etal00} with integrals over the whole volume,
 which define the $\lambda_i$ implicitly:

 \begin{equation}
 \int_V Y_i[\lambda_i^2 B_i-\alpha^2 B_i -(\nabla \alpha \times {\bf B}_i)] dV=0
 \label{eq41}
 \end{equation}

 This volume integration,
 which has to be carried out for every point in the volume
 is certainly very time consuming (a sixth order process).
 The $\lambda_i$ have the same dimension as
 the magnetic field. The existence of the $\lambda_i$ has been confirmed for the
 semi-analytic field of \cite{low:etal90} by \cite{li:etal04}.
 While the work of  \cite{li:etal04} showed that one can find the auxiliary
 function ${\bf {\bar Y}}$ for a given force-free field in 3D, the difficulty is
 that ${\bf \bar{Y}}$ is a-priori unknown if only the photospheric magnetic
 field vector is given. \cite{yan:etal00} proposed an iterative scheme to
 compute the auxiliary functions and the nonlinear force-free magnetic field
 selfconsistently. They use the approximate solution $k$ on the right hand
 side of Eq. (\ref{yan1}) to compute a better solution $k+1$ by:

 \begin{equation}
 c_i {\bf B_i}^{(k+1)}= \oint_S \left({\bf \bar{Y}^{(k)}} \frac{\partial {\bf B^{(k)}}}{\partial n}
 - \frac{\partial {\bf \bar{Y}^{(k)}}}{\partial n} {\bf B_0} \right) \; {\bf dS},
 \label{yan2}
 \end{equation}

 where the initial guess for the magnetic field in the volume is
 ${\bf B}=0$ and also the initial $\frac{\partial {\bf \bar{Y}}}{\partial n}=0$.
 In principle it would be also possible to compute a potential field first and
 derive the auxiliary functions for this field as done in
 \cite{li:etal04} and iterate subsequently for the nonlinear force-free fields
 and the associated auxiliary functions with Eq. (\ref{yan2}). This possibility
 has not been tried out to our knowledge until now, however.
 The method iterates the magnetic field until
 ${\bf B}$ and $\frac{\partial {\bf B}}{\partial n}$ converge.
 In an inter code comparison by \cite{schrijver:etal06} one
 iteration step of (\ref{yan2}) took about $80h$ for this method and only this one
 step was carried out without further iteration. This seems, however,
 not to be sufficient to derive an accurate nonlinear force-free solution.
 The method has been applied for the comparison
 with soft X-ray loops observed with YOHKOH by \cite{wang:etal00,liu:etal02}
 and to model a magnetic flux robe by \cite{yan:etal01,yan:etal01a}.

 In a new implementation of the boundary element
 method by \cite{yan:etal06} the auxiliary functions are computed iteratively
 with the help of a simplex method. This avoids the numerical expensive
 computation of the volume integral (\ref{eq41}).
 The boundary element method is still
 rather slow if a magnetic field has to be computed in an entire 3D domain.
 Different from other method, it allows, however, to evaluate the NLFFF-field
 at every arbitrary point within the domain from the boundary data, without
 the requirement to compute the field in an entire domain. This is in particular
 useful if one is interested to compute the NLFFF-field only along a given
 loop.

 \cite{he:etal06} investigated the validity of the boundary integral
 representation for a spherical implementation. The method has been tested
 with the longitudinal invariant \cite{low:etal90} solution. The spherical
 implementation method of this method revealed reasonable results for smooth
 modestly nonlinear fields, but a poor convergence for complex
 magnetic field structures and large values of $\alpha$.
\section{How to deal with non-force-free boundaries and noise ?}
Given arbitrary boundary conditions of the magnetic field vector on the
 photosphere, the solution to the force-free equations in 3D may not exist.
 Nonlinear force-free coronal magnetic field models assume, however, that
 the solution exists. It is certainly possible and necessary to check after
 or during the computation if a solution has been found. In the following
 we will discuss what we can do if the measured photospheric data are
 incompatible with the assumption of a force-free coronal magnetic field.
 \subsection{Consistency check of vector magnetograms}
 We reexamine some necessary conditions with the photospheric field
 (or bottom boundary of a computational box). These conditions have
 to be fulfilled in order to be suitable boundary conditions for a
 nonlinear force-free coronal magnetic field extrapolation.
 An a-priori assumption about the photospheric data is that the
 magnetic flux from the photosphere is sufficiently distant from the
 lateral boundaries of the observational domain and
 the net flux is in balance, i.e.,

 \begin{equation}
  \int_{S} B_z(x,y,0) \;dx\,dy =0.
 \end{equation}

 \cite{molodensky69,molodensky74,aly89,sakurai89} used the virial theorem to
 define which conditions a vector magnetogram has to fulfill to be consistent with the
 assumption of a force-free field in the corona above the boundary.
 These conditions are:

\begin{enumerate}
\item The total force on the boundary vanishes

 \begin{eqnarray}
 \int_{S} B_x B_z \;dx\,dy = \int_{S} B_y B_z \;dx\,dy = 0 &&
 \label{prepro1} \\
 \int_{S} (B_x^2 + B_y^2) \; dx\,dy  = \int_{S} B_z^2 \; dx\,dy. &&
 \label{prepro2}
 \end{eqnarray}

\item The total torque on the boundary vanishes

 \begin{eqnarray}
 \int_{S} x \; (B_x^2 + B_y^2) \; dx\,dy &=& \int_{S} x \; B_z^2 \; dx\,dy
 \label{prepro3} \\
 \int_{S} y \; (B_x^2 + B_y^2) \; dx\,dy &=& \int_{S} y \; B_z^2 \; dx\,dy
 \label{prepro4} \\
 \int_{S} y \; B_x B_z \; dx\,dy &=& \int_{S} x \; B_y B_z \; dx\,dy
 \label{prepro5}
 \end{eqnarray}

\end{enumerate}
 In an earlier review \cite{aly89} has mentioned already that the
 magnetic field is probably not force-free in the photosphere, where
 ${\bf B}$ is measured
  because the plasma $\beta$
 in the photosphere is of the order of one and pressure and gravity
 forces are not negligible. The integral relations
 (\ref{prepro1})-(\ref{prepro5}) are not satisfied in this case in
 the photosphere and the measured photospheric field is not a
 suitable boundary condition for a force-free extrapolation.
 \cite{metcalf:etal95} concluded that the
 solar magnetic field is not force-free in the photosphere, but becomes
 force-free only at about $400km$ above the photosphere. \cite{gary01} pointed
 out that care has to be taken when extrapolating the coronal magnetic
 field as a force-free field from photospheric measurements, because the
 force-free low corona is sandwiched between two regions (photosphere and
 higher corona) with a plasma $\beta \approx 1$, where the force-free assumption
 might break down. An additional problem is that measurements of the photospheric
 magnetic vector field
 contain inconsistencies and noise. In particular the transverse components
 (say $B_x$ and $B_y$) of current  vector magnetographs include uncertainties.

 The force-free field in a domain requires the Maxwell
 stress (\ref{prepro1})-(\ref{prepro5}) to sum to zero over the boundary.
 If these conditions are not fulfilled a force-free field cannot be found
 in the volume. A faithful algorithm should therefore have the capability
 of rejecting a prescription of the vector field at the boundary that fails
 to produce zero net Maxwell stress.  A simple way to incorporate these conditions
 would be to evaluate the integrals (\ref{prepro1})-(\ref{prepro5})
 within or prior to the NLFFF-computation and to refuse the vector field if the
 conditions are not fulfilled with sufficient accuracy. Current codes do run,
 however, although if feeded with inconsistent boundary data, but they certainly
 cannot find a force-free solution in this case (because it does not exist).
 This property of current codes does, however, not challenge the trustworthiness
 of the algorithms, because the force-free and solenoidal conditions are
 checked in 3D, e.g., with the help of the functional $L$ as defined in (\ref{defL}).
 A non zero value of $L$ (within numerical accuracy) tells the user that a
 force-free state has not been reached. In principle it would be possible
 that the codes do refuse to output the magnetic field in this case.
 For current codes this is not automatically controlled but responsibility
 of the user.

 Unfortunately current measurements of the magnetic field vector are only
 available routinely in the photosphere, where we have a finite $\beta$ plasma
 and non-magnetic forces might become important. The force-free compatibility
 conditions (\ref{prepro1})-(\ref{prepro5}) are not fulfilled in the photosphere,
 but they should be fulfilled in the low $\beta$ chromospheric and coronal plasma
 above. The question is if we still can use the photospheric measurements
 to find suitable consistent boundary conditions for a nonlinear force-free modelling.
 Such an approach has been called preprocessing of vector magnetograms.
\subsection{Preprocessing}
The preprocessing routine has been developed by
 \cite{wiegelmann:etal06}. The integral relations
 (\ref{prepro1})-(\ref{prepro5}) have been used to define
 a 2D functional of quadratic forms:

 \begin{equation}
  L_{\rm prep} = \mu_1 L_1 + \mu_2 L_2 + \mu_3 L_3 + \mu_4 L_4
 \end{equation}

where

\begin{eqnarray}
L_1 &=& \left[ \left(\sum_p B_x B_z \right)^2
              +\left(\sum_p B_y B_z \right)^2
  +\left(\sum_p B_z^2-B_x^2-B_y^2 \right)^2 \right]  \\
L_2 &=& \left[ \left(\sum_p x  \left(B_z^2-B_x^2-B_y^2 \right) \right)^2
              +\left(\sum_p y  \left(B_z^2-B_x^2-B_y^2 \right) \right)^2
        \right. \nonumber \\ & & \left. \hspace*{0.8em}
              +\left(\sum_p y B_x B_z -x B_y B_z \right)^2
        \right] \\
L_3 &=& \left[ \sum_p \left(B_x-B_{xobs} \right)^2
              +\sum_p \left(B_y-B_{yobs} \right)^2 \right. \nonumber \\
 &&      \left. +\sum_p \left(B_z-B_{zobs} \right)^2 \right] \\
L_4 &=& \left[ \sum_p \left(\Delta B_x \right)^2
                     +\left(\Delta B_y \right)^2
                     +\left(\Delta B_z \right)^2
        \right]
\end{eqnarray}

The surface integrals are here replaced by a summation $\sum_p$ over all grid nodes
$p$ of the bottom surface grid and the differentiation in the smoothing term is
achieved by the usual 5-point stencil for the 2D-Laplace operator. Each constraint
$L_n$ is weighted by a yet undetermined factor $\mu_n$. The first term ($n$=1)
corresponds to the force-balance conditions
 (\ref{prepro1})-(\ref{prepro2}), the next ($n$=2) to the torque-free condition
 (\ref{prepro3})-(\ref{prepro5}). The following term ($n$=3) ensures
that the optimized boundary condition agrees with the measured photospheric data and
the last terms ($n$=4) controls the smoothing. The 2D-Laplace operator is designated
by $\Delta$.
The aim of the preprocessing procedure is to minimize $L_{\rm prep}$ so that all
terms $L_n$ if possible are made small simultaneously. A strategy on how to find the
optimal yet undefined parameters $\mu_n$ is described in \cite{wiegelmann:etal06}.
As result of the preprocessing
 we get a data set which is consistent
 with the assumption of a force-free magnetic field in the corona but
also as close as possible to the measured data within the
 noise level.
\section{Code testing and code comparisons}
 Newly developed codes for the extrapolation of nonlinear force-free fields
 from boundary data have to be tested before they are applied to measurements.
 In principle any analytical or numerically created solution of the
 force-free equations  (\ref{forcebal})-(\ref{solenoidal}) can be used as a
 reference case. One cuts a plane (artificial photosphere, bottom
 boundary) out of the 3D reference solution
 \footnote{For the pure task of code testing it is also acceptable
 to use all six boundaries of the reference solution. These kind
 of data are not available for real solar cases of course.}
 and uses the
 above described extrapolation codes to reconstruct the magnetic field.
 The result of this extrapolation is then compared with the reference to
 rate the quality of the reconstruction. Unfortunately, it is very hard to
 find a truly nonlinear 3D solution of (\ref{forcebal})-(\ref{solenoidal})
 analytically and very few solutions are known. \citet{low:etal90} (LL)
 found a class of solutions which have become a standard reference for
 testing NLFFF-extrapolation codes.
 LL found  axisymmetric equilibria which are separable in
spherical coordinates. They are self-similar in the radial coordinate, and the polar
angle dependence is determined from a nonlinear eigenvalue equation. The symmetry is
broken by cutting out a rectangular chunk of the solution by using a Cartesian
coordinate system which is shifted  and rotated with respect to the original
coordinate system in which the LL equilibria are calculated. The parameters of the
LL solutions and the parameters of the new Cartesian coordinate system allow for a
large number of different situations which can be used for tests. The original
axisymmetric spherical LL solution has also been used (with and without symmetry
breaking by shifting the origin of the coordinate system) to test spherical
 NLFFF programs. To our knowledge all recent implementations of the described
 NLFFF approaches have been tested with LL, either immediately  in the
 original code-describing papers or in subsequent works, e.g., in
 a blind-algorithm test within the NLFFF-consortium, as described below.

 The MHD-relaxation method and the optimization approach have been compared
  by \cite{wiegelmann:etal03}. Both methods have been applied to the
  \cite{low:etal90} equilibrium with exactly the same finite difference grid.
  The iterative equations for MHD-relaxation and optimization
 have both the form
 $\frac{\partial {\bf B}}{\partial t}=\mu {\bf F}$
 but the structure of ${\bf F}$ is  more complicated for
 optimization than for MHD-relaxation. The MHD-relaxation term
 is indeed identical with the first term of the optimization approach.
 While MHD-relaxation minimizes only the Lorentz-force, the optimization
 does additional minimize $\nabla \cdot {\bf B}$, while a decreasing
 magnetic field divergence during MHD-relaxation (as seen in
 \cite{wiegelmann:etal03}) is the result of numerical diffusion. Despite
 the numerical overhead in computing ${\bf F}$ for the optimization code,
 optimization provided more accurate results and faster convergence.

 A practical advantage of the MHD-approach
 is that several time-dependent MHD-codes are well known and
 established and can be used for the force-free relaxation
 discussed here. The inclusion of non-magnetic forces like
 pressure gradients and gravity looks straight forward for
 the MHD approach.
 Other methods are usually developed with the only
 task of computing nonlinear force-free coronal magnetic fields, also
 a generalization towards magnetohydrostatic and stationary MHD-equilibria
 is possible and has been done for the optimization approach
 \citep[see][]{wiegelmann:etal03a,wiegelmann:etal06b}.
 Another advantage of using time-dependent MHD-codes for relaxation
 is that the computed force-free equilibrium can be used on the
 same grid and with the same code as initial state for time-dependent MHD-simulations.
 One can, in principle, use the force-free equilibria computed with any of the
 described method as initial state for time-dependent MHD-simulation, but
 having the initial equilibrium state already directly on the MHD-grid might be very handy,
 because no further adjustments are needed.
%
 \subsection{The NLFFF-consortium}
\begin{figure}
  \includegraphics[width=15cm,height=18cm]{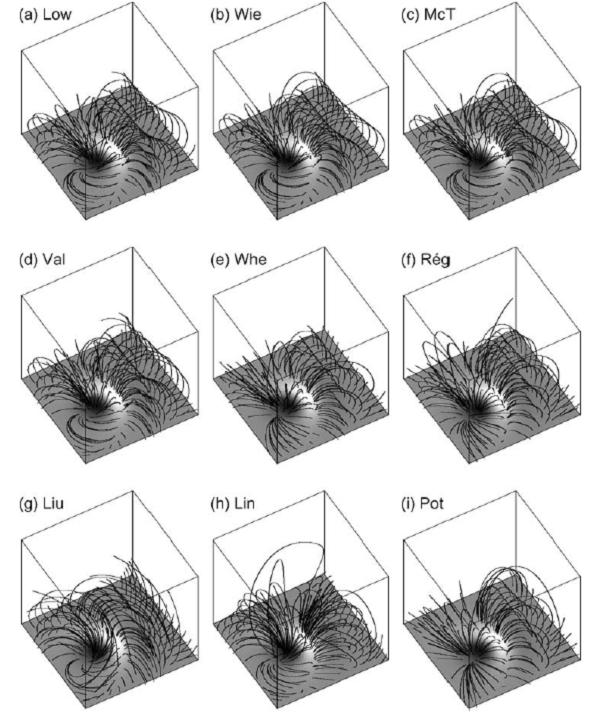}
  \caption{Evaluation of six nonlinear force-free codes. The reference solution (a)
  has been compared with extrapolations with 'optimization' (b,c);
  'MHD-relaxation' (d), 'Grad-Rubin' (e,f), 'Boundary element' (g).
  For comparison a linear-force-free (h) and potential field (i) are shown, too.
  The images show the central domain of the model. Only the bottom boundary has been
  used provided for the extrapolation.
 [This figure has been originally published as Fig. 4 in
 \cite{schrijver:etal06}. Used with permission of Springer.]}
  \label{figure3}
\end{figure}

 Since the year 2004 activities are ongoing to bring NLFFF-modelers together and
 to compare the different existing codes. A workshop series has been organized
 for this aim by Karel Schrijver and three workshops took place so far from
 2004-2006. The next workshop is planed for june 2007. As we have been asked
 to summarize the workshop results on the CSWM-meeting, we give also a very brief
 overview in the corresponding special issue-paper here. The main results of the first two
 workshops have been published in \cite{schrijver:etal06}.
 In this paper six different NLFFF-implementations
 (Grad-Rubin codes of \cite{amari:etal99} and \cite{wheatland04}, MHD-relaxation
 code of \cite{valori:etal05}, optimization codes by McTiernan and
 \cite{wiegelmann04}, boundary element method by \cite{yan:etal00}) have been compared.
 The codes have been tested in a blind algorithm test with the help of the
 semi-analytic equilibrium by \cite{low:etal90} in two cases. In case I
 all six boundaries of a computational
 box have been described and in case II only the bottom boundary. The comparison
 of the extrapolation results with the reference solution has been done qualitatively
 by magnetic field line plots
 (Shown here in Fig. \ref{figure3} for the central region of
 case II) and quantitatively by a number of sophisticated
 comparison metrices. All NLFFF-fields agreed best with the reference field for
 the low lying central magnetic field region, where the magnetic field and electric
 currents are strongest and the influence of the boundaries lowest.
 The code converged with speeds that differed by a factor of one million per
 iteration steps
\footnote{The codes run on different machines, have been written in different
programming languages and used different compilers. A real test of the exact
computing time would comprise a proper operation count, e.g., the number
of fixed point additions and multiplications per iteration step.}.
 The fastest-converging and best-performing code was the \cite{wheatland:etal00}
 optimization code as implemented by \cite{wiegelmann04}. Recent implementations
 of the Grad-Rubin code by \cite{amari:etal06,inhester:etal06} and a new
 implementation of the upward integration method by \cite{song:etal06} did not
 participate in the blind-algorithm inter-comparison by \cite{schrijver:etal06}, but
 these three new codes have been tested by the authors with similar measures and
 revealed similar accuracy as the best performing codes in the blind algorithm test.
 It seems that the somewhat more flexible boundary conditions used in the Grad-Rubin
 approaches of \cite{amari:etal06} and \cite{inhester:etal06} are responsible
 for the better performance compared to the earlier implementation by
 \cite{amari:etal99}, which has been used in the blind algorithm test.

 The widely used LL-equilibrium  contains a very smooth photospheric
 magnetic field and an extended
 current distribution. It is therefore also desirable to test NLFFF-codes
 also with other, more challenging boundary fields, which are less smooth, have
 localized current distribution and to investigate also the effects of
 noise and effects from non force-free boundaries. A somewhat more challenging
 reference case is the equilibrium found by \cite{titov:etal99} (TD).
 Similar as LL, the TD equilibrium is an axisymmetric equilibrium.
 The TD-model contains a potential field which is disturbed by a
 toroidal nonlinear force-free current. This equilibrium has been
 used for testing the MHD-relaxation code (Valori and Kliem,
 private communication) and the optimization code in \cite{wiegelmann:etal06a}.

 Any numerically created NLFFF-model might be suitable for code
 testing, too. It is in particular interesting to use models, which
 are partly related on observational data. Very recently
 \cite{vanballegooijen:etal07} used line-of-sight photospheric
 measurements from SOHO/MDI to compute a potential field, which
 was then disturbed by inserting a twisted flux robe and relaxed
 towards a nonlinear force-free state with a magnetofrictional
 method as described in \cite{vanballegooijen04}. The
 \cite{vanballegooijen:etal07} model is not force-free in
 the entire computational domain, but only above a certain
 height above the bottom boundary (artificial chromosphere).
 On the lowest boundary (photosphere) the model contains
 significant non-magnetic forces. Both the chromospheric
 as well as the photospheric magnetic field vector from the
 \cite{vanballegooijen:etal07}-model have been used to test
 four of the recently developed extrapolation codes
 (One Grad-Rubin method, one MHD-relaxation code and two
 optimization approaches)
 in a second blind algorithm test by \cite{metcalf:etal07}.
 While the I. NLFFF-consortium paper (\cite{schrijver:etal06}) used a domain
 of just $64^3$ pixel, the II. paper used a computational domain
 of $320 \times 320 \times 258$ pixel and modern NLFFF-codes
 where able to compute the nonlinear force-free field in such
 relatively large boxes within a few hours for a moderate
 parallelization on only 1-4 processors and a  memory requirement of
 $2.5-4$ GB of Ram. This very recent code-comparison shows a major
 improvement regarding computing time and suitable grid sizes within
 less than three years. On the first NLFFF-consortium
 meeting in 2004, box sizes of some $64^3$ have been a kind of standard
 or computing times of some two weeks have been reported for $150^3$ boxes.
 We briefly summarize the results of \cite{metcalf:etal07} as:
 \begin{itemize}
 \item NLFFF-extrapolations from chromospheric data recover the original
       reference field with high accuracy.
 \item When the extrapolations are applied to the photospheric data,
       the reference field is not well recovered.
 \item Preprocessing of the photospheric data improve the result, but
       the accuracy is still lower as for extrapolations from the chromosphere.
 \end{itemize}
\section{Conclusions and Outlook}
 Within the last few years the scientific community showed a growing
 interest into coronal magnetic fields. \footnote{Publications
 containing the phrase 'coronal magnetic fields' in title or abstract
 have been cited
 less than about $50$ times per year until the early 1990th and this number
 increased to about $150$ citations per year in 2004. A peak year was
 2006 (last year) with more than $300$ citations. (Source: ISI Web of Knowledge,
 march 2007)}
 The development of new ground based and space born vector magnetographs provide
 us measurements of the magnetic field vector on the suns photosphere. Accompanied
 from these hardware development, software has been developed to extrapolate the
 photospheric measurements into the corona. Special attention has recently been
 given to nonlinear force-free codes. Five different numerical approaches
 (Grad-Rubin, upward integration, MHD-relaxation, optimization, boundary elements)
 have been developed for this aim. It is remarkable that new codes or major
 updates of existing codes have been published for all five methods within the
 last two years, mainly in the last year (2006). A workshop series (NLFFF-consortium)
 since 2004 on nonlinear force-free fields has recently released synergy effects,
 by bringing modelers of the different numerical implementations together to
 compare, evaluate and improve the programs. Several of the most recent
 new codes and utility programs (e.g. preprocessing) have at least been partly
 inspired by these workshops. The new implementations have been tested with
 the smooth semi-analytic Low-Lou-equilibrium  and showed
 reasonable agreement with this reference field. While all methods aim for
 a reconstruction of the coronal magnetic field from the photospheric magnetic
 field vector, the way how these measurements are used to prescribe the
 boundaries of the codes is different.
 \begin{itemize}
 \item MHD-relaxation and optimization use  $B_{x0}, B_{y0}, B_{z0}$
 on the bottom boundary. This over-determines the boundary value
 problem. Both methods are closely related and compute
 the magnetic field in a computational box with

 \begin{equation}
 \frac{\partial {\bf B}}{\partial t}= \mu {\bf F},
 \end{equation}

 where the structure of {\bf F} is somewhat different (the optimization approach
 has more terms) for both methods. Usually a potential field is used as initial
 state for both approaches, also the use of a linear force-free initial state
 is possible. Recently a multiscale version of optimization has been
 installed, which uses a low resolution NLFFF-field as input for higher resolution
 computations. Specifying the entire magnetic field vector on the bottom
 boundary is an over-imposed problem and a unique
  NLFFF-field (or a solution at all) requires
 that the boundary data fulfill certain consistency criteria. A recently developed
 preprocessing-routine helps to find suitable consistent boundary data from
 inconsistent photospheric measurements.
 Earlier and current comparisons showed
 a somewhat higher accuracy for the optimization approach. A practical advantage
 of the MHD-approach is that in principle any available time-dependent MHD-code can be
 adjusted to compute the NLFFF-field.
 \item The Grad-Rubin approach uses  $B_{z0}$ and the distribution of
 $\alpha$ computed with Eq. (\ref{alpha0direct}) for one polarity, which
 corresponds to well posed mathematical problem. A practical problem is that
 the computation of $\alpha$ requires numerical differences of the noisy and forced
 transverse photospheric field  $B_{x0}, B_{y0}$ with (\ref{j_photo}) leading
 to inaccuracies in the normal electric current distribution and in $\alpha$.
 For smooth semi-analytic test cases this is certainly not a problem, but
 real data require special attention (smoothing, preprocessing, limiting
 $\alpha \not=0$ to regions where $B_{z0}$ is above a certain limit)
 to derive a meaningful distribution of $\alpha$. While the method
 requires only $\alpha$ for one polarity, the computation from
 photospheric data provide $\alpha$ for both polarities. We are not
 aware of any tests on how well NLFFF-solutions computed from $\alpha$
 prescribed on
 the positive and negative polarity coincide. It is also unclear how
 well the computed transverse field components on the bottom boundary
 agree with the measured values
 \footnote{In principle $B_{x0}, B_{y0}$ may have an additional field
 $(B_{x0}, B_{y0})+(\partial_x \partial_y) \phi$ without making a difference
 for $\alpha$ and hence for the Grad-Rubin result.}
  of $B_{x0}, B_{y0}$. More tests on
 this topics are necessary, including the recently installed possibility
 to prescribe $\alpha$ for both polarities and adjust the boundary by
 a weighed average of $\alpha$ on both polarities to fulfill
 Eq. (\ref{alphaeq}). As initial state
 the Grad-Rubin method uses a potential field, which is also true
 for MHD-relaxation and optimization.
 \item The upward integration and the boundary element method prescribe
 both all components of the bottom boundary magnetic field vector and
 the $\alpha$ distribution computed with Eq. (\ref{alpha0direct}).
 This approach over imposes the boundary and $B_{x0}, B_{y0}, B_{z0}$
 and $\alpha$ have to be consistent which each other and the
 force-free assumption. This is certainly
 not a problem at all for smooth semi-analytic test equilibria and
 strategies to derive consistent boundary data from measured data
 have been developed recently. Different from the three approaches
 discussed above, upward integration and boundary element methods
 do not require to compute first an initial potential field in the
 computational domain.
 It is well known that the upward integration method is based on an
 ill-posed problem and the method has not been considered for
 several years, but a recent implementation with smooth analytic functions
 might help to regularize this method. First tests showed a reasonable results
 for computations with the smooth semi-analytic Low-Lou solution.

 The boundary element method has the problem to be very slow and an earlier
 implementation of this method could not reach a converged state for
 a $64^3$ boxed used in the I. NLFFF-consortium paper due to this problem.
 A new 'direct boundary method' has been developed, which seems to be faster
 than the original 'boundary element method',
 but still slower compared with the four other NLFFF-approaches if
 the task is to compute a 3D magnetic field in an entire 3D-domain.
 Different from all other described methods the boundary element
 approach allows to compute the nonlinear force-free field vector at
 any arbitrary point above the
 boundary and it is not necessary to compute the entire 3D-field above
 the photosphere. This might be a very useful feature if one is
 interested in computing the magnetic field only along a single loop and
 not interested in an entire active region.

 The new implementations of upward integration and boundary element method
 show both reasonable results for first tests with the smooth semi-analytic
 Low and Lou equilibrium.
 Further tests with more sophisticated equilibria, e.g. a solar-like
 test case as used in the II. NLFFF-consortium
 paper would be useful to come to more sound conclusions regarding the
 feasibility of these methods.
 \end{itemize}
 Most of the efforts done in nonlinear force-free modelling until
 now concentrated mainly on developing these models and testing
 their accuracy and speed with the help of well known test configuration.
 Not too many applications of nonlinear force-free models to real data are
 currently available, from which we learned new physics. One reason was the
 insufficient access to high accuracy photospheric vector magnetograms and
 a second one were limitations of the models. Force-free field extrapolation
 is a mere tool, if properly employed on vector magnetograms, it can help to
 understand physical, magnetic field dominated processes in the corona.
 Both the computational methods as well as the
 accuracy of required measurements (e.g. with Hinode, SDO) are rapidly
 improving. Within the NLFFF-consortium we just started (since april 2007)
 to apply the different codes to compute nonlinear force-free coronal
 magnetic fields from Hinode vectormagnetograms. This project might provide
 us already some new insights about coronal physics.

 To conclude, we can say that the capability of Cartesian nonlinear force-free
 extrapolation codes has rapidly increased in recent years.
 Only three years ago most codes run usually on grids of about $64^3$ pixel.
 Recently developed or updated codes (Grad-Rubin by Wheatland,
 MHD-relaxation by Valori, optimization by Wiegelmann, optimization by
 McTiernan) have been applied to grids of about $300^3$ pixel.
 Although this increase of traceable grid sizes is certainly encouraging,
 the resolution of current and near future vector magnetographs
 (which of course measure only data in 2D!) is significantly higher.
 We should keep in mind, however, that the
 currently implemented NLFFF-codes have been only moderately parallelized
 using only a few processors. The CSWM-conference, where this paper has been
 presented, took place at the 'Earth simulator' in Yokohama, which contains
 several thousands of processors used for Earth-science computer simulations.
 An installation of NLFFF-codes on such massive parallel computers
 (which has been briefly addressed on NLFFF-consortium meetings) combined
 with adaptive mesh refinements might enable drastically improved grid sizes.
 One should not underestimate the time and effort necessary to program
 and install such massive parallelized versions of existing codes.
 As full disk vectormagnetograms will become available soon (SOLIS, SDO/HMI)
 it is also an important task to take
 a spherical geometry into account. First
 steps in this direction have been carried out with the optimization and
 boundary element methods. Spherical NLFFF-geometries are currently still
 in it's infancy and have been tested until now only with smooth semi-analytic
 Low and Lou equilibria and require further developments.

 Attention has also recently been drawn to the problem that
 the coronal magnetic field is force-free, but the photospheric
 one is not. Tests with extrapolations from solar-like
 artificial photospheric and chromospheric measurements within
 the II. NLFFF-consortium paper revealed that extrapolations from
 the (force-free) chromospheric field provide significantly better
 results as extrapolations using directly the (forced) photospheric
 field. Applying a pre-processing program on the photospheric data,
 which effectively removes the non-magnetic forces, leads to significantly
 better results, but they are not as good as by using the chromospheric
 magnetic field vector as boundary condition. An area of current research
 is the possibility to use chromospheric images to improve the preprocessing
 of photospheric magnetic field measurements. Improvements in measuring the
 chromospheric magnetic field directly \citep[e.g.][]{lagg:etal04}
 might further improve to find suitable boundary conditions
 for NLFFF-extrapolations. Force-free extrapolations are not suitable,
 however, to understand the details of physical processes on how the
 magnetic field evolves from the forced photosphere into the chromosphere,
 because non-magnetic forces are important in the photosphere.
 For a better understanding of these phenomena more sophisticated models
 which take pressure gradients and gravity (and maybe also plasma flow) into account
 are required. Some first steps have been done with a generalization of
 the optimization method by \cite{wiegelmann:etal06b}, but such approaches
 are still in their infancy and have been tested so far only with smooth
 MHD-equilibria. It is also not entirely clear how well necessary
 information regarding the plasma (density, pressure, temperature, flow)
 can be derived from measurements. Non-magnetic forces become important also
 in quiet sun regions (\cite{schrijver:etal05}) and in
 the higher layers of the corona, where the plasma $\beta$ is of the order of unity.
 Coronagraph measurements, preferably from two viewpoints as
 provided by the STEREO-mission, combined with a tomographic inversion
 might help here to get insights in the required 3D structure of the
 plasma density. One should also pay attention to the combination of extrapolation
 methods, as described here, with measurements of the Hanle and Zeeman
 effects in coronal lines which allows the reconstruction of the
 coronal magnetic field as proposed in feasibility studies of vector
 tomography by \cite{kramar:etal06,kramar:etal07}. Other measurements of
 coronal features, e.g., coronal plasma images from two STEREO-viewpoints,
 can be used for observational tests of coronal magnetic field models.
 Using two  viewpoints provide a much more restrictive test of models as
 images from only one view direction. While a nonlinear force-free
 coronal magnetic field model helps us to derive the topology, magnetic field
 and electric current strength in coronal loops, they do not provide
 plasma parameters. One way to get insights regarding the coronal plasma is the use
 of scaling laws to model the plasma along the reconstructed 3D field lines and
 compare correspondent artificial plasma images with real coronal images.
 \cite{schrijver:etal04} applied such an approach to global potential coronal
 magnetic fields and compared simulated and real coronal images from one viewpoint.
 A generalization of
 such methods towards the use of more sophisticated magnetic field models
 and coronal images from two STEREO-viewpoints will probably provide many insights
 regarding the structure and physics of the coronal plasma. An important challenge
 is for example the coronal heating problem. The dominating coronal magnetic field
 is assumed to play an important role here, because magnetic field configuration
 containing free energy can under certain circumstances reconnect
 (\cite{priest96,priest99}) and supply energy for coronal heating.
 \cite{priest:etal05} pointed out that
 magnetic reconnection at separators and separatrices plays an important role
 for coronal heating. Nonlinear force-free models can help here to identify
 the magnetic field topology, magnetic null points, separatrices and
 localized strong current concentration. While magnetic reconnection
 \citep[see e.g.][]{priest:etal99} is a dynamical phenomenon,
 the static magnetic field models discussed here can help to identify
 the locations favourable for reconnection. Time sequences of
 nonlinear force-free models computed from corresponding vector magnetograms
 will also tell wether the topology of the coronal magnetic field has changed due
 to reconnection, even if the physics of reconnection is not described by
 force-free models. Sophisticated 3D coronal magnetic field models and
 plasma images from two viewpoints might help to constrain the coronal heating
 function further, which has been done so far with plasma images from one viewpoint
 \citep[][by using data from Yokoh, Soho and Trace]{aschwanden01,aschwanden01a}.
%
%
%
%
\begin{acknowledgments}
I would like to thank the organizers of CSWM for inviting me to give a
review talk on nonlinear force-free magnetic field modelling. The author
thanks Bernd Inhester for useful comments and acknowledge inspiriting
 discussions with many other NLFFF-modelers on three workshops organized by Karel Schrijver
 (NLFFF-consortium) in Palo Alto between 2004 and 2006 and at various other occasions.
 This work was supported  by DLR-grant 50 OC 0501.
\end{acknowledgments}
  %

%

\end{article}
\end{document}